# BLOCKREDUCE:
# SCALING BLOCKCHAIN TO HUMAN COMMERCE

**Karl J. Kreder III**
Chief Security Officer, GridPlus Inc
Instructor, The University of Texas at Austin
karl@gridplus.io

## ABSTRACT

Blockchains have shown great promise as peer-to-peer digital currency systems over the past 10 years. However, with increased popularity, the demand for processing transactions has also grown, leading to increased costs and confirmation times as well as limited blockchain utility. There have been a number of proposals on how to scale blockchains, such as Plasma, Polkadot, Elastico, RapidChain, Bitcoin-NG, and OmniLedger. These solutions all propose the segmentation of every function of a blockchain, namely consensus, permanent data storage, transaction processing, and consistency, which significantly increases the complexity and difficulty of implementation. BlockReduce is a new blockchain structure which only segments consistency, allowing it to scale to handle tens of thousands of transactions per second without impacting fault tolerance or decentralization. Moreover, BlockReduce will significantly decrease node bandwidth requirements and network latency through incentives while simultaneously minimizing other resource demands in order to prevent centralization of nodes.

***K*eywords** blockchain · scaling · Bitcoin · BlockReduce · proof-of-work

## 1 Introduction

Blockchains have the potential to disrupt the way that humans organize and collaborate in groups. Presently, once a collaborative group reaches a certain size, hierarchies are introduced as the mechanism for scaling collaborative environments with growing group sizes. Representatives then speak for a subset of people within the larger group to allow the group to grow in size and tackle problems of ever growing complexity. However, the hierarchical structure of collaborative organization lends itself to representatives within the hierarchy taking advantage of the position and the people they represent. Blockchains promise to introduce a new system of human organization that enables individuals to collaborate in a voluntary, peer-to-peer manner without power-based hierarchies. Unfortunately, as blockchains have gained adoption, current distributed blockchain systems are encountering severe limitations of scaling beyond ~10 transactions-per-second (TPS). Currently proposed solutions to the scaling problem introduce some form of centralization, points of trust, or authority in a hierarchy. These proposed solutions, although potentially better than the current hierarchical organizational systems, would likely devolve into new forms of pseudo-decentralized, dysfunctional hierarchies.

BlockReduce presents a new blockchain topology that offers 3+ orders of magnitude improvement in transaction throughput while avoiding the introduction of hierarchical power structures and centralization. This is accomplished through a modification to the block reward incentive to not only reward work, but to also reward optimization of network constraints and efficient propagation of transactions. This is accomplished by creating Proof-of-Work (PoW) managed hierarchy of tightly-coupled, merge-mined blockchains.

## 2 Scaling requirements

Blockchains are primarily useful for applications that involve the transfer of value between two or more distinct economic parties [1]. Although there may exist some use cases for blockchains without transfer of value, these applications can likely be fulfilled by simpler, more efficient technologies such as event-sourced databases. Multiple blockchain scaling proposals have been made in literature. However, prior proposals have been challenged in delivering scalability without sacrificing one or more principal characteristics which allow blockchains to facilitate value transfer. Therefore, the intrinsic characteristics of a blockchain must first be defined and scaling requirements be clarified so that differing scaling proposals can be compared and assessed on their individual merits and drawbacks.

New architectures which enable higher performance should not decrease the usefulness or utility of the blockchain for value transfers. Although there are many characteristics which define blockchains, the ones which are most pertinent to scaling are the throughput (transactions per second), verifiability, Byzantine fault tolerance, fungibility, and economic sustainability. Throughput is relatively easy to define as the number of transactions per second (TPS) that a blockchain can process. For discussion we can establish a target throughput of 50,000 TPS, which is roughly 3 times the average rate of consumer transactions worldwide in 2015 [2, 3]. Verifiability is the ability for blockchain participants to provably verify the current chain state as well as the state transitions creating the current state. Byzantine fault tolerance is the most difficult property to understand when considering blockchains because many factors play into assessing a blockchain's fault tolerance [4, 5]. These include, but are not limited to, economic decentralization of nodes, the percentage of the network's hash power which must act nefariously to violate the consensus rules, and the economic cost of obtaining that hash power. Byzantine fault tolerance can be assumed to decrease if PoW hash power is split via sharding or if a validating authority or committee is introduced [6]. Centralization decreases Byzantine fault tolerance, because it decreases the number of economic actors in the system who must collude to attack the chain. Fungibility in the context of scaling is having a system in which the unit of account has nearly equivalent economic value regardless of "where" an asset exists. Put simply, if the state of a blockchain is sharded, the free market exchange rate of the unit of account from one shard to another should be as close to parity as possible. Finally, there is the issue of economic sustainability. Economic sustainability is the ability of the scaling solution to generate sufficient revenue to guarantee the security of the blockchain over the long term.

## 3 Previous Work

Previous proposals for scaling blockchain have met some of the requirements proposed in the previous section, but none of them have satisfied all requirements. For example, Plasma [7] falls short because it reduces Byzantine fault tolerance, will likely create price disparities across various plasma chains, and does not have an enforceable mechanism to pass revenue back to the main chain. The Lightning Network [8] for Bitcoin is a layer-2 solution that allows significant increases in performance through cascaded payment channels. Unfortunately, Lightning introduces hubs which must provide collateral to pass through payments, causing centralization and capital-based transaction costs. This decreases the network fault tolerance while also making the system less sustainable, because fees are kept by hub operators rather than going to secure the blockchain. Proposals such as Bitcoin-NG, Elixxer, OmniLedger, RapidChain, and Elastico, all introduce some form of committee and shard work with the introduction of sharding of state [9, 10, 11, 12, 13]. These designs decrease the fault tolerance of the system and could introduce fungibility problems with uncontrolled state sharding. FruitChain [14] has proposed an interesting mechanism in which incremental PoW is used to aggregate transactions without sharding of state. This generally meets all of the requirements except for performance and sustainability, due to throughput limitations. Binary blockchains [15] have also been proposed, which allow higher throughput by sharding state between a number of parallel blockchains. However, this decreases the Byzantine fault tolerance of the chain by splitting hash power between the parallel chains. BlockReduce combines the ideas of incremental work and sharding of state with merge-mining to form a tightly coupled PoW-managed hierarchy of blockchains which satisfies all of the proposed scaling requirements.

## 4 BlockReduce

### 4.1 Consistency

The definition of consistency is key to the idea of BlockReduce and the ability to scale blockchains. The following is a brief definition of what is encompassed by the idea of consistency.

consistency - The state where nodes agree on a set of transactions on which to mine.

It is necessary to reiterate that consistency is not consensus. Consistency can be reached within a subset of the overall network participants on a subset of pending transactions without having reached consensus. The subset of participants may agree on what transaction set they are working to validate and publish, but the transactions within that set are not globally consistent. Transactions only become globally consistent once a certain amount of work is performed and consensus is reached. Therefore, even though transactions are described as being consistent, they are not yet confirmed or settled until consensus is reached. Another way to understand the difference between consistency and consensus is that consistency is localized consensus and consensus is global consistency.

### 4.2 Proof-of-Work Hierarchy

The introduction of a hierarchy in distributed systems that do not lend themselves to centralization is technically challenging. Most proposals for systems that have a form of hierarchy attempt to address this, although inadequately, through the use of randomization. Unlike other proposals, BlockReduce creates an auto-balancing hierarchy of self-identifying groups which are formed to efficiently reach consistency. BlockReduce is also unique in that consensus is still reached globally. The PoW hierarchy is only a mechanism of efficient co-operation for data propagation. To enable scalability that could encompass all human commerce, BlockReduce establishes a three-level hierarchy as shown in Figure 2. Each square in Figure 2 represents a self-contained blockchain and the blockchains are linked in the hierarchy by three mechanisms. The hashes of the previously-mined lower level blocks are recorded in the higher level blocks. Transactions which exceed the scope of a blockchain are passed to the higher levels until the required scope is attained. Finally, the header of the blockchains allow the three levels of blockchains to be simultaneously merge-mined. Merge-mining will be discussed in greater detail later.

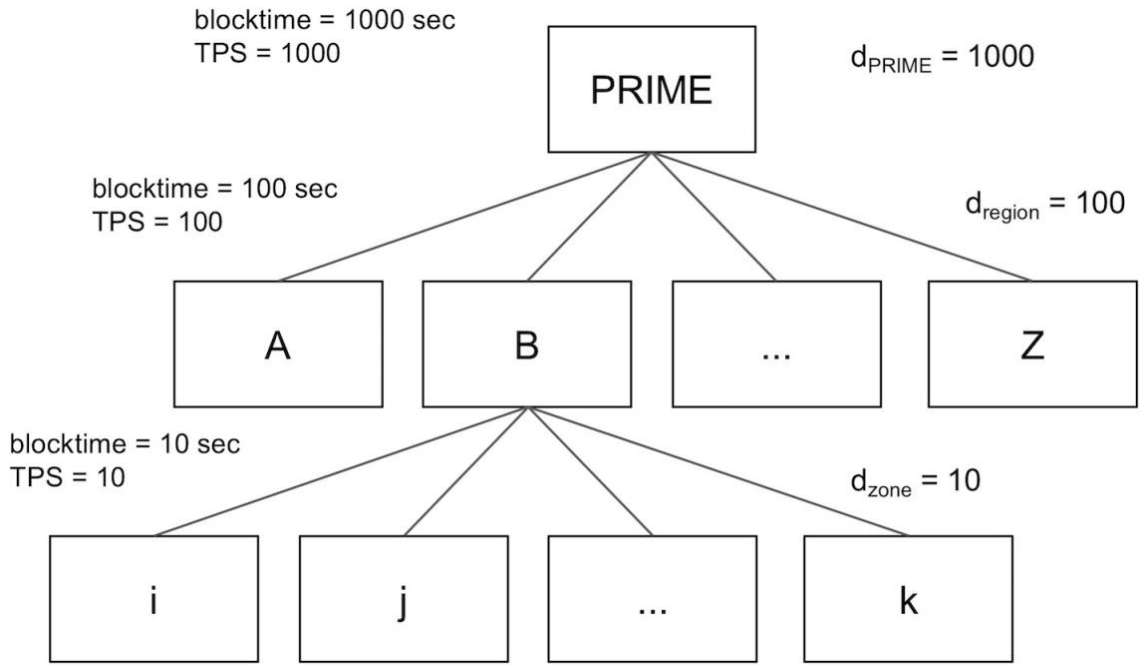

Figure 1: proposed BlockReduce hierarchy, illustrating groups and sub-groups.

The root blockchain of this hierarchy is referred to as PRIME. All nodes work to validate all transactions included in PRIME. Additionally, nodes perform work, transaction verification, and data propagation in one of the subsequent levels referred to as regions (A-Z) and one of the subordinate zones (i-k). The lowest level, zones such as i and j, are assumed to be peer-to-peer groups communicating via a gossip protocol. Therefore, each one can be thought of in terms of size and throughput as a traditional blockchain such as Bitcoin or Ethereum. We assume that each zone is similar in performance to an Ethereum blockchain, which roughly has 10 second blocks, approximate throughput of 10 transactions per second (TPS), and a basic transaction size of 100 bytes. Based on these assumptions as well as a hierarchy that has 10 regions and 10 zones in each region, a BlockReduce type blockchain would be able to process 1,000 TPS. The PRIME block time will average 1,000 seconds (~17 minutes) and the total chain will require around 8 Tb/year of storage. Most importantly, the amount of bandwidth used to create 1,000 TPS blocks is significantly smaller than that of current blockchains. Additionally, the structure of blockchains can be dynamically scaled over time by altering the number of valid regions and zones. The re-sizing of a structure to 70 regions and 70 zones in each region would allow a throughput of at least 50,000 TPS. Moreover, the segmentation of the network into zones will likely result in geographically organized peer groups, optimized for network latency. This efficient grouping of peers will achieve higher throughput requiring fewer regions and zones.

### 4.3 Segmentation and Aggregation

Currently, transactions are shared and propagated across most blockchain networks through a gossip protocol. Nodes select an arbitrary number of peers within the network and share newly received transactions with their selected peers. Each transaction that a node receives is broadcast to all peers even if they have already received the transaction. This creates a large network bandwidth requirement simply for the network to propagate transactions. The inefficiency in bandwidth usage is due to the fact that a transaction has to be transmitted for a node to determine if it is new. Therefore, if a node already has seen a transaction, the full bandwidth must be expended to determine that the transaction does not need to be transmitted. Transactions must be propagated for the nodes to create a consistent set of transactions to validate and include in the next block. In Bitcoin, the bandwidth overhead for a node compared to the amount of data actually included in a block is on the order of 100-300x. This means that for every 1 Mb of data included in a block, a node needs to send and receive 100-300 Mb of data [16]. The bandwidth overhead requirement can be reduced if information can be grouped into larger sets. This can allow peers to exchange identifying information for a data set, say a hash or header, before the peers actually have to transfer the entire data set.

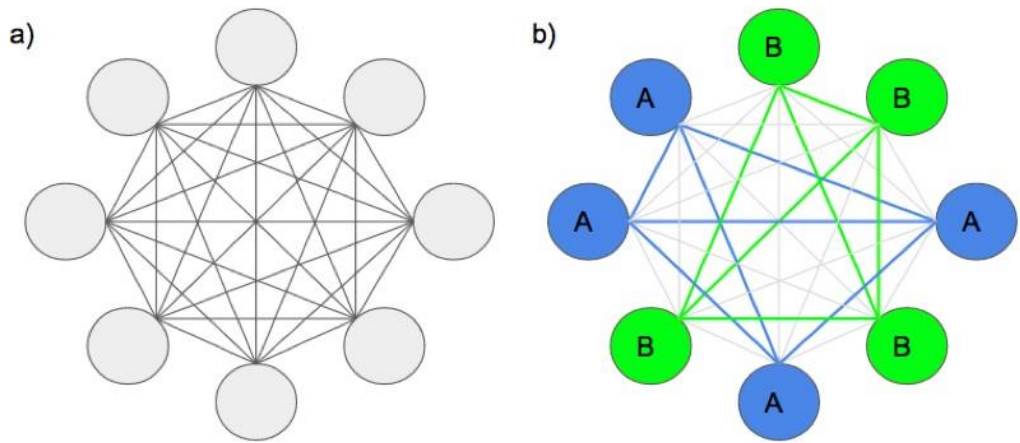

Figure 2: a) traditional peer-to-peer network b) self-imposed subdivision of peer-to-peer network.

Table 1: Bitcoin vs. BlockReduce requirements with increasing block size

| Characteristic | Scale Factor | Block Size (PRIME) (MB) | | | | | | |
|---|---|---|---|---|---|---|---|---|
| | | 2.5 | 10 | 20 | 40 | 80 | 160 | 320 |
| Transaction per Second (TPS) | N | 10 | 40 | 80 | 160 | 320 | 640 | 1280 |
| Transactions per Block | N | 10 | 40 | 80 | 160 | 320 | 640 | 1280 |
| Storage per year (Gb) | N | 75 | 300 | 600 | 1200 | 2400 | 2800 | 9600 |
| Average Bandwidth (kB/s) | N | 250 | 1000 | 2000 | 4000 | 8000 | 160000 | 32000 |
| Yearly Traffic (Tb) | N | 7.3 | 29.4 | 58.7 | 117.5 | 235 | 670 | 940 |
| BR Average Bandwidth (kb/s) | k + N/100 | 250 | 760 | 770 | 790 | 830 | 910 | 1070 |
| BR Yearly Traffic (Tb) | k + N/100 | 7.3 | 22.3 | 22.6 | 23.2 | 24.4 | 26.7 | 31.4 |
| BR Bandwidth Comparison (%) | n/a | 100% | 76% | 38.5% | 19.8% | 10.4% | 5.7% | 3.3% |

Functionally, this means that if a mechanism can be constructed that incentivizes peers to share information with a subset of their peers as blocks of data rather than individual transactions, a significant increase in bandwidth efficiency can be achieved. For example, Figure 2(a) illustrates a traditional peer-to-peer architecture currently used by blockchain systems. Since each node does not know what transactions have been received by its peers, it must pass every new transaction it receives to each of its peers. However, if the network is subdivided into sub-groups, as in figure 2(b), the peers would only share the individual transactions with other peers within their sub-group. Specifically, nodes which self-identify as A would only share single transactions with other nodes which also self-identify as A. The same would also be true for nodes which self-identify as sub-group B. At some point, once a set of transactions is consistent within sub-group A, sub-group A would share that set of transactions with sub-group B. Now that the transactions are built up into larger groups of data, say hundreds or thousands of transactions, nodes can first check if they need to transmit data to their peers by checking a hash. This enables a significant increase in bandwidth efficiency. Transmitting a hash would only take tens of bytes, but would allow the nodes to determine if they need to transmit several kilobytes of transaction data. The next question that arises is how would nodes self-organize and then how do zones determine when to share data with other zones. Table 1 shows a comparison on the bandwidth and storage requirements for a Bitcoin style blockchain versus a BlockReduce style blockchain for various block sizes. The most notable item is that the scale factor for BlockReduce bandwidth is $k+N/100$, compared to $N$ for Bitcoin. The large reduction in the BlockReduce bandwidth scale factor is what enables BlockReduce to massively scale.

### 4.4 Merge-Mining and Difficulty

A key aspect of BlockReduce is all nodes will mine at the zone, region, and PRIME level at the same time. The simultaneous mining of PRIME, regions, and zones allows BlockReduce to keep the entire network's proof-of-work on PRIME. Incremental work via region and zone blocks also allows work to act as a mechanism to determine consistency within regions and zones. Simultaneous mining is accomplished using the well-known idea of merge-mining. When a node is mining a transaction, it will include not only a block reward transaction for the PRIME block but will also include a block-reward transaction for the region and zone block. The difficulty of the region and zones are set such that many zone blocks will be found inside of each region block and many region blocks will be found inside each PRIME block. Difficulty for region blocks is set to some fraction of the PRIME difficulty and zone blocks to some fraction of the region difficulty. For example, as illustrated in Figure 2, the difficulty of PRIME is set to 1,000 and the region and zone difficulties are 100, and 10, respectively. This creates a system where ~10 zone blocks will be found for each region block and ~10 region blocks will be found for every PRIME block. The ratio of difficulty from one level to the next in the hierarchy does not necessarily need to be 10, but it is used here for illustrative purposes.

The meaning of a region block and zone block need to be defined and their purpose explained. A region block and zone block do not imply that consensus has been reached, they only imply that local consistency has been reached. Once local consistency is reached, the region or zone block can be shared with other peer-level regions or zones. This delay in propagation of transactions only after aggregation is the fundamental mechanism that allows more efficient bandwidth utilization. Using the assumptions laid out in Figure 2, up to 100 transactions will be included in a zone block running at or near capacity. Once transactions are formed into a zone block, only the header of the zone block is broadcast to region peers. Only if a peer determines a block is new do they request the entire zone block. This means that ~1/100th of the data contained in a zone block is transmitted, which is far more efficient than transmitting every transaction in the zone block many times. Even greater efficiency is achieved with region blocks. If there are 1,000 transactions in the average region block, again only the header, which represents 1/1,000th of the data in a region block, needs to be transmitted between peers before sharing a region block. The ability to propagate data as blocks, by request, removes significant bandwidth requirements as the number of TPS increases.

### 4.5 Transaction Structure and Scope

Additional fields need to be added to a typical Bitcoin transaction, which indicate the destination location of the transaction. Specifically, a 2 byte field indicating the region 0-255 and the zone 0-255 need to be added to the transaction header. If multiple inputs are used in a transaction, they must reside within the same scope, meaning inputs must be taken from the same zone. A UTXO can be spent to any other zone or region within the network by signing a new transaction that relocates it to a new zone. The cost of making the transaction as well as the time to settle will be determined by the scope of the transaction. If a transaction moves a token from one zone and spends it in the same zone, it would be a zonal transaction and would settle by the next zone block. If the transaction moves a token from one zone to another, but the zones have a common parent region, then it would be a regional transaction. This would cost slightly more, and settle when the next region block is found. If a transaction is made from one zone to another which does not have a common region parent, it would cost the most, and would only settle when the next PRIME block is found. These transactions would have zonal, regional, and PRIME scope, respectively. Note, that when a transaction is constructed and broadcast, it must be broadcast and processed by nodes which are mining the zone in which it is located.

The idea of location is essentially changing the scope of a UTXO. If a user desires to change the scope of the UTXO, they can do so by paying more to process a regional or PRIME transaction. There are two benefits to locating UTXOs. First, locating transactions is an effective mechanism for making SPAM attacks (broadcasting a conflicting transaction to more than one zone) impossible. Second, it allows local consistency to become global consensus for transactions which don't modify scope. This allows transactions the option to be able to fully settle at both the region and zone blocks in addition to PRIME. By restricting the region and zone that can process located UTXOs, the local consistency that is reached when the corresponding region or zone blocks are found now becomes global consensus. The time and cost for higher scoped transactions will be higher. The cost of a PRIME transaction to the network is higher in terms of bandwidth, computing, and permanent data storage compared to a zonal transaction and the transaction fee should necessarily reflect the higher cost.

The transaction location requirement for BlockReduce will incentivize users to self-separate into roughly equally utilized zones. Users will seek the greatest amount of utility (i.e. fast, cheap transactions) with the largest number of likely counter parties. The transaction throughput of each zone will roughly reach a state of equilibrium. This will cause higher transaction fees for a busier zone and incentivize movement of transactions to less busy zones over time. Interestingly, the zones which users aggregate into will be based on a combination of geography, network topology, and existing economic relationships. When this is accounted for, the zones may be thought of as regional currencies. This is very similar to how major fiat currencies, such as the US dollar, Euro, or Yen are segmented, traded, and used today. The main difference is that funds in each zone will maintain free market value parity due to the mining and transaction incentive structures.

### 4.6 Block Header Structure

In the proposed block structure, a zone block, a region block, and a PRIME block's data look apparently identical. This allows for merge-mining to be used to mine zone, region, and PRIME blocks simultaneously. The characteristic that distinguishes a zone, region, and PRIME block is the level of difficulty achieved when a block is mined. The difficulty of the zone will be roughly one order of magnitude easier than a region block which will be roughly one order of magnitude easier than a PRIME block. Consequently, some of the data that is recorded in a zone block is not useful in the canonical record.

Table 2: A proposed block header structure to implement BlockReduce

| Field Size | Name | Data Type | Description |
| --- | --- | --- | --- |
| 4 | Version | uint32 | Block version information |
| 32 | parentPrime | char[32] | The hash value of the previous parent block |
| 32 | parentRegion | char[32] | The hash value of the previous region block |
| 32 | parentZone | char[32] | The hash value of the previous zone block |
| 32 | merkleRootPrime | char[32] | Merkle root of all PRIME transactions |
| 32 | merkleRootRegion | char[32] | Merkle root of all region transactions |
| 32 | merkleRootZone | char[32] | Merkle root of all zone transactions |
| 32 | merkleRootInterlink | char[32] | Merkle root interlink enabling proof compression |
| 4 | unixTime | uint32 | Current time in seconds |
| 4 | bitsPrime | uint32 | Current PRIME difficulty in compact form |
| 4 | bitsRegion | uint32 | Current region difficulty in compact form |
| 4 | bitsZone | uint32 | Current zone difficulty in compact form |
| 8 | feesRegion | uint64 | Total transaction fees included in region block |
| 8 | feesZone | uint64 | Total transaction fees included in a zone block |
| 1 | mapRegion | uint8 | Current valid number of regions |
| 1 | mapZone | uint8 | Current valid number of zones |
| 1 | locationRegion | uint8 | Region where block is mined |
| 1 | locationZone | uint8 | Zone where block is mined |
| 8 | nonce | uint64 | Block nonce |

Specifically, if a zone block is found, any PRIME Merkle Root Hash (MRH), PRIME TX, region MRH, and region TX are data which have not yet reached consensus. Therefore, the record of that data will be different in both order in structure relative to the data which is ultimately recorded in the canonical ledger. However, the recorded hashes are not completely wasted. They provide a mechanism to validate data sets shared with peer regions and peer zones when a region or zone block is found. When a zone block is found, the zone block containing zone MRH as well as data which has equivalent or higher scope will be shared with peer zones. This includes transactions which have region or PRIME scope which are represented by the region MRH and the PRIME MRH included in the zone block. The peer zones will then check the shared data (i.e. region and PRIME TX's) and append it to their working block.

### 4.7 Block Incentives

The network incentives need to be modified to incentivize the regions and zones needed for BlockReduce. In the most abstract form, the incentive structure is changed from rewarding only mining to rewarding efficient group formation and efficient, low-latency data propagation. The region and zone blocks are the mechanism by which rewards for these new non-PoW driven goals are measured. An initial proposal of the new block reward distribution scheme is summarized as follows:

$$R = P_{reward}(1 - \chi) + P_{reward} \left[ \frac{b_r}{d_{PRIME}/d_{group}} + \frac{b_z}{d_{PRIME}/d_{zone}} \right] \chi \qquad (1)$$

where $P_{reward}$ is a constant representing the target reward amount, $\chi$ is a weighting between rewarding PoW and network latency, $d_{PRIME}$ is the difficulty of a PRIME block, $d_{region}$ is the difficulty of a region block, $d_{zone}$ is the difficulty of a zone block, $b_g$ is the number of region blocks, and $b_z$ is the number of zone blocks. The target award amount, $P_{reward}$ for a given block is the sum of the newly minted coins, as dictated by monetary policy, and the transaction fees of all transactions included in the PRIME block. The actual reward amount, $R$, could vary slightly from the target reward amount owing to the stochastic variation in the actual number of region and zone blocks found. This difference can be made up in the short term by adjusting the number of newly minted coins. Moreover, in the long term, the impact of the variations on monetary policy would be negligible. However, at the point at which monetary policy dictates a near zero expansion in the monetary base, the transaction fees will be the majority of the reward. The ability to adjust the actual reward, given that the target reward is composed of transaction fees, will be limited due to the fact that coins cannot be artificially destroyed. At this end-point operating case, any excess transaction fees could be sequestered into an escrow account which can then be drawn down when the actual reward exceeds the target reward. Any shortfall in the escrow account can be made up by the minting of new coins. This would effectively hold the inflation rate to zero while still maintaining a varying incentive needed to encourage the even distribution of hash power among the zones.

### 4.8 Auto-Load Balancing

The block reward mechanism is designed to incentivize nodes and miners to self-organize into efficient regions and zones. The impact of latency on mining, which manifests as uncle blocks, fosters efficient self-organization within the context of the new block reward construct. A miner is penalized by having a network connection and a group of peers which causes that miner to experience high network latency. This is manifested by the fact that both transactions and blocks take longer to propagate. The delay in propagation means that transactions are less likely to be included in a block, causing a decrease in earned transaction fees. Moreover, a delay in receiving a mined block increases the likelihood of working on an uncle block and decreases the amount of time for which hash power is effectively used. Unfortunately, in the context of a single group of network peers and block rewards which are largely composed of PoW reward, nodes are both limited in their ability and incentive to address the network issues. However, once the nodes are provided the option of selecting a group, they will do so in a way which minimizes the network latency. This means that groups will form which are geographically close, or more precisely close in terms of network topology and latency. The incentive to find geographically close groups will be counterbalanced by the incentive to find groups which have the least amount of mining power. As seen by the proposed block reward mechanism, nodes are rewarded proportionally to the number of region and zone blocks that they find. Therefore, they will try to form groups where the network latency incentive is balanced with the desire to find regions and zones with the lowest difficulty.

The current architecture also incentivizes users of the network to use the network in an efficient way. When a user creates a transaction, they will need to specify a zone to process the given transaction. This ensures that the transaction is only processed and originated by a single zone. Additionally, a user will pick a zone which is the least busy because they will be able to get their transaction processed more quickly for a lower fee. A user will be able to pick the least busy zone by either performing an analysis of recent transactions on chain or consuming data from an analytics service provider. The proposed reward mechanism, coupled with each user's desire to seek the least expensive and most expeditious transaction processing, will cause the network participants to auto-load balance the network on both the supply and demand side simultaneously.

### 4.9 Dynamic Network Sizing

The network will be able to automatically adjust the number of regions and zones to dynamically match transaction processing supply to current demand. The BlockReduce protocol will set a range of valid regions and zones within a maximum range of 256. Nodes and miners choose which regions and zones to operate within. Based on the current demand, as determined by the last N blocks in PRIME, the protocol will re-adjust the range of valid regions and zones to modulate total supply. The BlockReduce protocol will target a zone block fill of say 80%. If the actual number of transactions significantly exceeds the 80% target for a significant period of time, the protocol will adjust the number of regions and zones to effectively increase transaction supply. If the block utilization falls below the target, the protocol will adjust to decrease supply. By maintaining a utilization near the target value, the protocol attempts to maximize total revenue from transaction fees. The maximization of transaction fee revenue is extremely important in the long run because transaction fees must pay for the PoW and network computing resources once the inflationary block rewards go to zero. Presently, all blockchains and blockchain scaling solutions are not economically viable in the long term for this reason. Current blockchains are not able to generate sufficient transaction fee revenue to pay for the network security. Blockchains must become sustainable through on-chain scaling because it is the only mechanism by which a sustainable cost per transaction can be realized in the long term.

There are some slight complications that need to be addressed with network re-sizing. Namely that there needs to be a mechanism by which the maximum range of valid regions and zones are deterministically re-mapped to the current valid range. This needs to be done to prevent transactions from becoming invalid when a re-sizing function occurs. Furthermore, it allows transactions to be formed and broadcast without requiring knowledge of the present state of the network.

### 4.10 Data Sharding

Locating UTXOs allows further sharding of the node functions, specifically transaction processing as well as persistent storage. Transactions that are processed at the zone level do not need to be recorded at the region or PRIME levels. Rather, only when a transaction exceeds zone scope does it need to be recorded at a higher level. Transactions which are scoped at the zone or region level can be recorded in the next higher level by simply recording the hash of child blocks in the parent. This allows a significant amount of bandwidth savings as well as reduction in the storage required for persistent data. Specifically, rather than needing some 8 Tb of storage per node per year, each zone node would only require 20 Gb of storage per year.

### 4.11 51% Attack Resistance

In the protocol where data sharding is allowed, the ability of the chain to deal with 51% attacks must be discussed. In the example in which there are 10 regions and 10 zones in each region, each zone will only have 1% of the total hash power of the network. If an attacker was able to achieve 0.6% of the total hash power and nefariously direct it within a single zone, they would represent 60% of the zone hash power and be able to pass nefarious zone block hashes into region blocks. However, the protocol already has a mechanism by which this type of attack will self-correct. If 60% of a zone attempts to pass a nefarious block into the region, the honest 40% of the zone hash-power will fork with the malicious 60%. The zone will begin publishing blocks at the region level that have significantly lower total hash-power than the other zones. Nodes in the other zones will be incentivized to move some of their hash power from their current zone to the one with less effective hash power. As new honest nodes move to the zone, they will decide to mine the honest fork with 40% of the hash power. Over time, the fork with 40% will garner more hash power and will become the canonical record of transactions. Furthermore, the incentive exists for nodes in other zones to monitor topologically adjacent zones to determine when a fork-induced arbitrage opportunity exists. It is not a requirement for zones to monitor adjacent zones, however validating adjacent zone blocks significantly de-risks nodes from working on incorrect blocks. Even if a zone node does not want to persist state or validate transactions from adjacent zone blocks, a node has some assurance on the block validity due to the work performed on the block. The more hash power that a node controls, the greater the economic incentive will be for them to validate more transactions and maintain a greater amount of state. The incentive of the miners to auto balance load across the zones and regions will work to mitigate 51% reversal attacks within zones on locally consistent transactions.

## 5 Conclusion

Scaling blockchains to handle a large number of transactions without compromising the inherent characteristics which allow them to function as money is a critical problem. BlockReduce proposes a novel PoW managed hierarchy of highly structured, merge-mined chains which can efficiently scale blockchain TPS by more than 3 orders of magnitude. This enables a performant blockchain which is also verifiable, fault tolerant, and economically sustainable. Interestingly, the structure proposed here is a direct acyclic graph which uses incentives to maintain a high degree of symmetry. Although it is possible to use other graph topologies, symmetry is likely a requirement to maintain fungibility.